\documentclass[twocolumn,showpacs,superscriptaddress,amsmath,amssymb]{revtex4}

\usepackage{graphicx}
\usepackage{dcolumn}
\usepackage{bm}

\begin{document}

\title{Direct observation of the superconducting gap in phonon spectra}

\author{F. Weber}
\altaffiliation[present address: ]{Materials Science Division, Argonne National Laboratory, Argonne, Illinois, 60439, USA.}
\affiliation{Forschungszentrum Karlsruhe, Institut f\"ur Festk\"orperphysik, P.O.B. 3640, D-76021 Karlsruhe, Germany}
\email{fweber@anl.gov}
\affiliation{Physikalisches Institut, Universit\"at Karlsruhe, D-76128 Karlsruhe, Germany}
\author{A. Kreyssig}
\affiliation{Technische Universit\"at Dresden, Institut f\"ur Festk\"orperphysik, D-01062 Dresden, Germany}
\author{L. Pintschovius}
\author{R. Heid}
\author{W. Reichardt}
\affiliation{Forschungszentrum Karlsruhe, Institut f\"ur Festk\"orperphysik, P.O.B. 3640, D-76021 Karlsruhe, Germany}
\author{D. Reznik}
\affiliation{Forschungszentrum Karlsruhe, Institut f\"ur Festk\"orperphysik, P.O.B. 3640, D-76021 Karlsruhe, Germany}
\affiliation{Laboratoire L$\acute{e}$on Brillouin, CE-Saclay, F-91911 Gif-sur-Yvette, France}
\author{O. Stockert}
\affiliation{Max-Planck-Institut f\"ur Chemische Physik fester Stoffe, D-01187 Dresden, Germany}
\author{K. Hradil}
\affiliation{Universit\"at G\"ottingen, Institut f\"ur Physikalische Chemie, D-37077 G\"ottingen, Germany}

%

%

\date{\today}

\begin{abstract}
We show that the superconducting energy gap $\Delta$ can be directly observed in phonon spectra, as predicted by recent theories. In addition, since each phonon probes the gap on only a small part of the Fermi surface, the gap anisotropy can be studied in detail. Our neutron scattering investigation of the anisotropic conventional superconductor YNi$_2$B$_2$C demonstrates this new application of phonon spectroscopy.
\end{abstract}

\maketitle

In their seminal paper\cite{Axe73}, Axe and Shirane found that the
onset of sueprconductivity in Nb$_3$Sn affects the frequencies and
the linewidths of phonons whose energies are comparable to the
superconducting energy gap $2\Delta$. Similar effects were
observed by neutron and Raman scattering in a number of
conventional superconductors\cite{Shapiro75}, but also in
high-$T_c$ copper oxides\cite{Krantz88,Friedl90,Pyka93,Limonov98,Strohm98}. In many
of these cases the authors argued that it was possible to
determine whether the phonon energy was above or below $2\Delta$
from the comparison of phonon frequences and linewidths above and
below $T_c$. Here we show that in the case of strong
electron-phonon coupling, the entire phonon lineshape can change
dramatically accross $T_c$
allowing a precise determination of $2\Delta$. 

A few years ago, strong superconductivity-induced changes of the
phonon lineshape were reported for a particular phonon in
YNi$_2$B$_2$C\cite{Kawano96} and LuNi$_2$B$_2$C\cite{Stassis97},
which - for the first time - could not be described by a simple
change in the phonon linewidth. This observation stimulated
further theoretical work\cite{Allen97,Kee97}. We performed very
precise inelastic neutron scattering measurements of many
different phonons in YNi$_2$B$_2$C ($T_c=15.2\,\rm{K}$) with a
much smaller statistical error covering a much larger region of
reciprocal space than in Ref. \onlinecite{Kawano96}.

The neutron scattering experiments were performed on the 1T
triple-axis spectrometer at the ORPHEE reactor at LLB, Saclay, and
on the PUMA triple-axis spectrometer at the research reactor FRM
II in Munich. Double focusing pyrolytic graphite monochromators
and analysers were employed in both cases. A fixed analyser energy
of $14.7\,\rm{meV}$ allowed us to use a graphite filter in the
scattered beam to suppress higher orders. The experimental
resolution was obtained from calculations using a formalism
checked on phonons with very low coupling strength. The wave
vectors are given in reciprocal lattice units of ($2\pi/a$
$2\pi/b$ $2\pi/c$), where $a = b = 3.51\,$\AA$\,$ and $c =
10.53\,$\AA. The single crystal sample of YNi$_2$B$_2$C weighing
2.26 g was grown by the floating zone method using the $^{11}$B
isotope to avoid strong neutron absorption. The single crystal
sample was mounted in a standard orange cryostat at LLB  and in a
closed-cycle refrigerator at FRM II, allowing us measurements down
to $T=1.6\,\rm{K}$ and $3\,\rm{K}$, respectively.

\begin{figure}
\begin{center}
\includegraphics[width=0.9\linewidth]{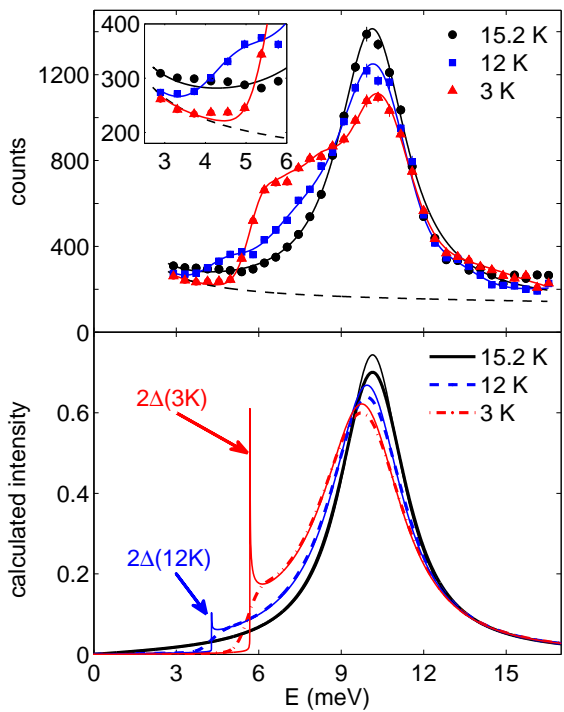}
\caption{\label{fig_1} (color online) \textit{top}: Evolution of the
neutron-scattering profile measured on YNi$_2$B$_2$C at \textbf{Q}
= (0.5,0.5,7) at $T=T_c(15.2\,\rm{K}$) and below $T_c$. Solid
lines are guides to the eye. The broken line denotes the
background as determined from measurements at neighboring
$\mathbf{q}$-points. \textit{bottom}: Calculated phonon lineshapes
based on the theory of Allen et al.\cite{Allen97}, using
parameters extracted from the experimentally observed lineshape in
the normal state (thin lines). Thick lines are obtained after
convolution with the experimental energy resolution.}
\end{center}
\end{figure}

Figure \ref{fig_1} shows the evolution through $T_c$ of the low
temperature lineshape of the phonon at the endpoint of the
transverse acoustic branch in the (110) direction. This phonon has
never been investigated before but its strong coupling to
electrons has been predicted by density functional theory\cite{Reichardt05}. It appears at \textbf{q} = (0.5, 0.5, 0) (the
M-point). On cooling from $T=200\,\rm{K}$ to a temperature just
above $T_c=15.2\,\rm{K}$, this phonon softens by about 10 \% and
broadens substantially, indicative of a strong electron-phonon
coupling (data not shown). However, the lineshape remains
Lorentzian to a very good approximation\footnote{The theoretical
lineshape is not a true Lorentzian, but is very close to a
Lorentzian for the case of the M-point phonon.}. On further
cooling through $T_c$, the lineshape starts to deviate strongly
from a Lorentzian. In particular, a step-like increase at a
certain energy $E_s$ appears as the temperature is lowered through
$T_c$ (fig.~\ref{fig_1}, top). $E_s$ increases with decreasing
temperature below $T_c$. According to the theory developed by
Allen et al. \cite{Allen97}, a part of the low energy tail which
lies below the gap, is pushed up in energy to form a narrow spike
at $2\Delta$(fig.~\ref{fig_1}, bottom). The theory is based on the
full quantum mechanical treatment of electron-phonon coupling
where vibrational and electronic excitations mix into hybrid
modes. The finite spectrometer resolution should wash out the
spike in experiments resulting in a sharp intensity increase at
$2\Delta$. The predicted intensity increase corresponds to the one
we observe at $E_s$. We show below that other experimental
features are also in an excellent agreement with this scenario.
Thus we will interpret our results using the theoretical framework
developed in Ref. \onlinecite{Allen97}.

\begin{figure}
\begin{center}
\includegraphics[width=0.99\linewidth]{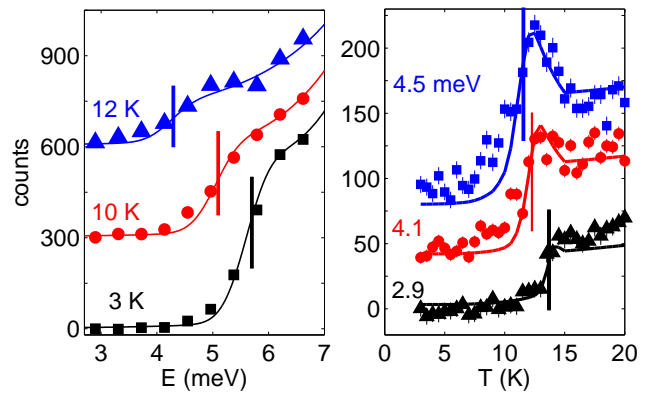}
\caption{\label{fig_2} (color online) \textit{left}: Background subtracted energy
scans taken at \textbf{Q} = (0.5,0.5,7) at three different
temperatures including an offset of 300 counts for clarity. Lines
were calculated from the theory of Allen et al.\cite{Allen97}
after scaling to the experimentally observed step-heights and
folding with the experimental energy resolution. The vertical bars
indicate the gap values used in the calculations, respectively.
\textit{right}: Temperature scans taken at \textbf{Q} =
(0.5,0.5,7) for three different energy transfers including an
offset of 35 counts for clarity. Lines are as in the left panel.
Vertical bars indicate the temperature for which the energy
transfer is equal to the gap value $2\Delta$ in the calculation.}
\end{center}
\end{figure}

A close comparison of the calculated lineshape with the low energy
part of the observed phonon spectral function can give highly
accurate values of $2\Delta$ as a function of temperature
(fig.~\ref{fig_2}, left). Close to $T_c$, where the step height is
small, it is more convenient to scan the temperature at a fixed
energy transfer: a sudden intensity decrease on cooling will mark
the temperature where the gap is
equal to the energy transfer (fig.~\ref{fig_2}, right). 

The theory of Allen et al.\cite{Allen97} predicts not only the
suppression of the spectral weight below $2\Delta$ just mentioned,
but also a downward shift of the intensity maximum and an
intensity build-up on the low energy side of the peak
(fig.~\ref{fig_1}, bottom). The observed changes are qualitatively
similar, but are somewhat stronger than predicted. These
discrepancies cannot be remedied by tuning the three parameters of
the calculations. Two of them, the phonon energy and the phonon
linewidth observed just above $T_c$, are entirely fixed by the
phonon properties in the normal state and the third, the gap value
is practically fixed as well by the sharp intensity increase at
$E_s$. Notwithstanding the shortcomings of the actual theory -
which might be due to the weak coupling approximation -, it
predicts the observed changes with temperature over a wide range
of energies remarkably well (fig.~\ref{fig_2}).

The features in the intensity vs. temperature curves
(fig.~\ref{fig_2}, right) related to the gradual opening of the
gap are quite sharp and, therefore, the gap values extracted from
the data depend very little on fine details of the model. The
temperature dependent gap value deduced in this way is plotted in
fig.~\ref{fig_4}. The experimental curve deviates somewhat from
the BCS weak coupling curve, which is not surprising for a
superconductor with a $T_c$ as high as $15.2\,\rm{K}$. Likewise,
the value of the zero temperature gap indicates strong coupling
because the ratio $2\Delta/k_BT_c=4.3$ ($k_B$ is the Boltzmann
constant) exceeds
considerably the weak coupling limit of $3.52$.

The method of extracting the superconducting gap from phonon
lineshapes explained above looks straightforward. Therefore, one
might ask why it has not yet been used so far. There are two
stringent conditions for making the replica of the superconducting
gap clearly visible: the phonon energy has to be considerably
larger than $2\Delta$ but on the other hand, the linewidth in the
normal state has to be large enough that the tail of the phonon
line extending into the gap region can still be discriminated from
the background. One might also think to study phonons with an
energy comparable to the gap energy. Here, the
superconductivity-induced changes of the phonon lineshape are
fairly strong, but extraction of the gap energy is possible only
with the help of detailed calculations. An example for such a case
is the acoustic phonon in the (100)-direction at wave vector
\textbf{q} = (0.5, 0, 0). Profound changes of the lineshape of
this phonon were first reported by Kawano et al.\cite{Kawano96}.
Puzzled by the appearance of an intense and relatively narrow line
below $T_c$ Kawano et al. thought to have found a new sharp
excitation. We reinvestigated this phonon and show that the
peculiar lineshapes appearing below $T_c$ can be quite naturally
understood using again the theory of Allen et al.\cite{Allen97}.

\begin{figure}
\includegraphics[width=0.9\linewidth]{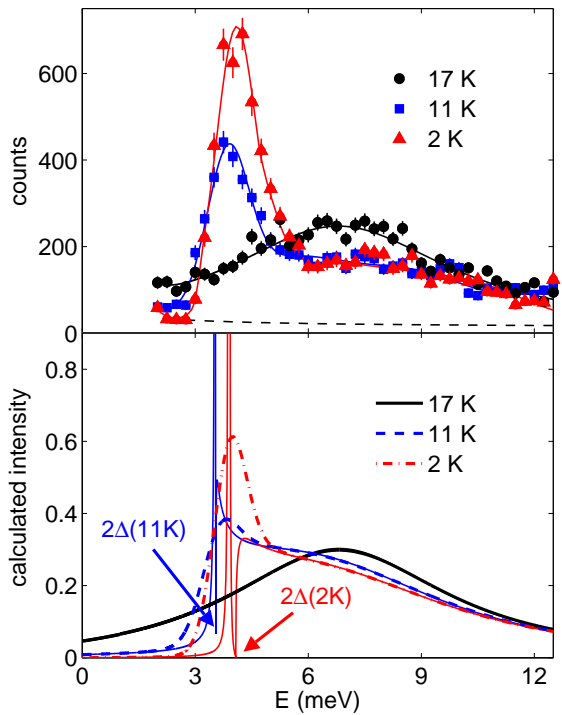}
\caption{\label{fig_3} (color online) \textit{top}: Evolution of the neutron-scattering profile measured on YNi$_2$B$_2$C at \textbf{Q} = (0.5,0,8) above and below $T_c$. Solid lines are guides to the eye. The broken line denotes the background as determined from measurements at neighboring $\mathbf{q}$-points. \textit{bottom}: Calculated phonon lineshapes based on the theory of Allen et al.\cite{Allen97}, using parameters extracted from the lineshape observed in the normal state (thin lines). The thick lines are obtained after convolution with the experimental resolution.}
\end{figure}
The phonon at \textbf{q} = (0.5, 0, 0) has a significantly lower
energy than the M-point phonon discussed above ($7\,\rm{meV}$ as
compared to $10\,\rm{meV}$) and moreover, has a much larger
linewidth\footnote{The lineshape of this phonon just
above $T_c$ is less well described by a Lorentzian because its
linewidth is closer to the phonon energy.}. As a consequence, the
superconductivity-induced redistribution of spectral weight is
even stronger, but the superconducting gap cannot be inferred from
the data as easily, except for temperatures close to $T_c$ where
the gap is still small. For lower temperatures, most of the
spectral weight condenses into a fairly sharp peak whose energy is
somewhat below the gap. According to Ref. \onlinecite{Allen97}, this
resonance can be regarded as a mixed vibrational/superelectronic
collective excitation. Its position with respect to $2\Delta$ has
to be determined from calculations for the parameter values of
this particular phonon. Again, the theory does not reproduce the
observed lineshapes quantitatively, but sufficiently well to
precisely determine the gap value (fig.~\ref{fig_3}).

\begin{figure}
  \begin{center}
    \includegraphics[width=0.9\linewidth]{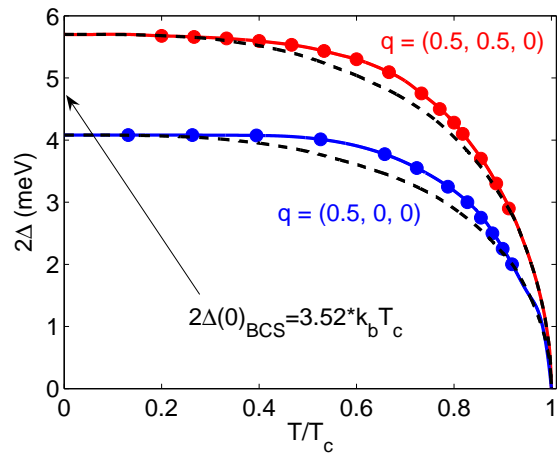}
    \caption{\label{fig_4} (color online) Superconducting energy gap $2\Delta$ versus temperature as evaluated from phonon data. Broken lines are calculated from weak-coupling BCS theory scaled to the experimental values at $T=3\,\rm{K}$. Red and blue points were deduced from phonon measurements at different momentum transfers. Lines are guides to the eye.}
  \end{center}
\end{figure}
Previous publications\cite{Axe73,Shapiro75} observed line
narrowing for phonons with an energy smaller than 2$\Delta$. We
found that this temperature dependence can be quantitatively
accounted for by the theory of Allen et al.\cite{Allen97}.
Unfortunately, there are no phonons in YNi$_2$B$_2$C with energies
$<$ 2$\Delta$, which have a sufficiently strong coupling to search
for temperature effects of the linewidth in this compound.

\begin{figure}
\begin{center}
\includegraphics[width=0.95\linewidth]{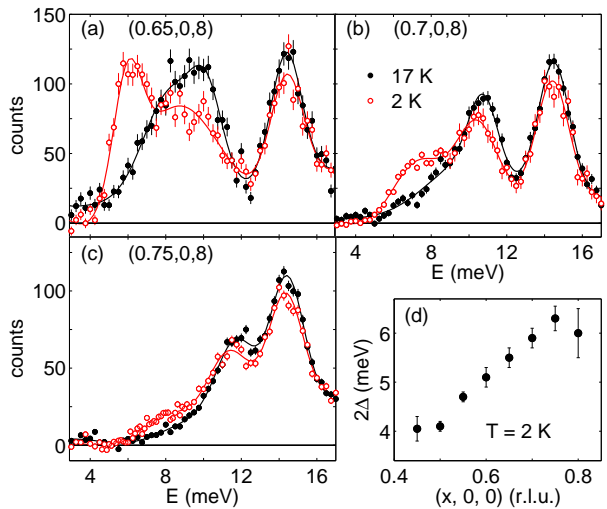}
\caption{\label{fig_5} (color online) Panels (a)-(c) show background subtracted energy scans taken at \textbf{Q} = (0.65-0.75,0,8) at $T=2\,\rm{K}$ (open symbols) and $17\,\rm{K}$ (filled symbols). Lines are guides to the eye. Panel (d) describes the evolution of the low temperature superconducting gap value extracted from phonon scans along the (100) direction.}
\end{center}
\end{figure}

The low temperature lineshapes of the two phonons discussed so
far look quite distinct. A study of several phonons along the
(100)-direction reveals that the lineshape may take on any
intermediate form, depending on the phonon energy relative to the
gap and on the normal state phonon linewidth. When going from
\textbf{q} = (0.5, 0, 0) towards the zone boundary, the normal
state phonon energy increases and the linewidth shrinks. As a
result, the low temperature lineshape gradually evolves from that
observed at \textbf{q} = (0.5, 0, 0) to that of the M-point phonon
(fig.~\ref{fig_5}, (a)-(c)). Fig.~\ref{fig_5} is interesting also
for another reason: the superconductivity-induced redistribution
of spectral weight involves the optic line at $14\,\rm{meV}$ as
well with the spectral weight near both peaks transfered to much
lower energies near the gap. At these \textbf{q}-points, the
acoustic and the optic modes strongly hybridize. Understanding
this phenomenon in detail would require a highly nontrivial
extension of the theory, which has not yet been made.

When plotting $2\Delta$ extracted from the various phonons studied
by us, it becomes clear that they probe different energy gaps,
ranging from $4\,\rm{meV}$ to more than $6\,\rm{meV}$ (figs. 4,
5(d)). Phonons probe the gap on the parts of the Fermi surface
connected by the phonon wave vector. Only theory can tell which
parts are actually involved. Density functional theory
calculations may provide such information, but they require the
technically involved solution of the anisotropic gap equations
including the full momentum dependence of the electron-phonon
coupling, which has not yet been carried out for YNi$_2$B$_2$C. In
special cases, however, the respective parts of the Fermi surface
can be inferred from the Fermi surface topology: for instance, the
extremely large linewidth of the \textbf{q} = (0.5, 0, 0) phonon
can be attributed to a nesting phenomenon predicted by
theory\cite{Reichardt05}.  \\
Whereas the intensity step observed at \textbf{q} = (0.5, 0.5, 0)
is very sharp, i.e. resolution-limited ($\approx1\,\rm{meV}$), it
is significantly broader ($\approx2\,\rm{meV}$) at \textbf{q} =
(0.7, 0, 0) and (0.75, 0, 0), and therefore cannot be linked to a
single, well-defined gap value. In this case the phonon wave
vector may connect several different pieces of the Fermi surface
probing a superposition of gap values.

Our results for the gap anisotropy in YNi$_2$B$_2$C fit nicely to
those extracted from tunnelling data using point-contact
spectroscopy\cite{Bashlakov07}. Further, the lowest gap seen in
our phonon experiments agrees very well with the value derived
from critical field $H_{c2}(T)$ data\cite{Shulga98}. On the other
hand, we have no evidence that the gap goes below
$2\Delta=4\,\rm{meV}$ anywhere on the Fermi surface and possibly
even to zero, as was proposed on the basis of
field-angle-dependent heat capacity measurements\cite{Park03} and
field-angle dependent thermal conductivity data\cite{Izawa02}.
Extensive energy and wave vector resolved density functional
theory calculations\cite{Reichardt05}, which are in very good
agreement with experiment\cite{Weber09}, did not show any other
suitable phonons in order to further probe the possible presence
of nodes.

To conclude, we confirm the prediction of Allen et al. that the
phonon line may acquire very peculiar shapes with prominent
features directly related to the superconducting energy gap
$2\Delta$. In contrast to a theoretical proposition\cite{Kee97},
this effect is not restricted to phonons with wave vectors close
to an extremum vector of the Fermi surface but may appear
throughout the Brillouin zone. The gap value $2\Delta$ can be
directly inferred from the lineshape, often without recourse to a
fit. This allows extraction of the gap values as a function of
temperature with high accuracy. By studying phonons of different
wave vectors, even the gap anisotropy can be explored by phonon
spectroscopy. Thus phonon spectroscopy can be a much richer source
of information on the superconducting properties than hitherto
assumed.


\end{document}